\documentclass[11pt,a4paper]{article}

\usepackage[
  top=26mm,
  headheight=12pt,
  headsep=5.15mm,
  text={160mm,216mm},
  marginparsep=5mm,
  marginparwidth=12mm,
  bindingoffset=6mm,
  footskip=10.13mm
]{geometry}
\usepackage{graphicx}
\usepackage{authblk}
\usepackage{amsmath,amssymb,amsfonts}
\usepackage{siunitx}
\usepackage{braket}
\usepackage{overpic}
\usepackage{bm}
\usepackage{xcolor}
\usepackage[hidelinks,colorlinks=true,linkcolor=blue,citecolor=blue,urlcolor=blue]{hyperref}
\usepackage[all]{hypcap} 
\usepackage[style=phys, backend = biber, biblabel=brackets, pageranges=false, url=false, doi=false, isbn=false, eprint=false]{biblatex}
\usepackage[capitalise]{cleveref}

\newcommand*\abs[1]{\lvert #1 \rvert}
\newcommand*\circled[1]{{\large \textcircled{\small #1}}}
\newcommand*{\sfref}[2]{Figure~\hyperref[#1]{\ref*{#1}#2}}
\AtEveryBibitem{\clearfield{url}}
\AtEveryBibitem{\clearfield{doi}}

\title{Talking with a ghost: semi-virtual coupled levitated oscillators}

\author{Ronghao Yin\thanks{These authors contributed equally to this work.}}
\author{Yugang Ren\protect\footnotemark[\value{footnote}]}
\author{Deok Young Seo}
\author{Anoushka Sinha}
\author{Jonathan D. Pritchett}
\author{Qiongyuan Wu}
\author{James Millen\thanks{Correspondence: \href{mailto:james.millen@kcl.ac.uk}{james.millen@kcl.ac.uk}}}

\affil{Department of Physics, King's College London, Strand, London, WC2R 2LS, UK}

\date{\today}

\begin{document}
\maketitle

\begin{abstract}
Mesoscopic particles levitated by optical, electrical or magnetic fields act as mechanical oscillators with a range of surprising properties, such as tuneable oscillation frequencies, access to rotational motion, and remarkable quality factors. Coupled levitated particles display rich dynamics and non-reciprocal interactions, with applications in sensing and the exploration of non-equilibrium and quantum physics. In this work, we present a single levitated particle displaying coupled-oscillator dynamics by generating an interaction with a virtual or ``ghost'' particle. This ghost levitated particle is simulated on an analogue computer, and its properties can thus be dynamically varied. Our work represents a new angle on measurement-based bath engineering and physical simulation and, in the future, could lead to the generation of novel cooling mechanisms and complex physical simulation.
\end{abstract}


\section{Introduction}
Levitated nano- and microparticles are oscillators with remarkable properties~\cite{Millen2020rev}, including ultra-high quality-factor oscillations at room temperature~\cite{Pontin2020, dania2024}, tuneable resonant modes, and~access to nonlinear rotational motion~\cite{Ahn2018}. The~ability to precisely monitor and control the motion of levitated objects enables precision force sensing~\cite{ranjit2016,Ahn2020, Liang2022, yin2022} and cooling to the quantum ground state of optical potentials~\cite{delic2020, Magrini2021, Tebbenjohanns2021, piotrowski2023}, with~a wide range of applications in fundamental physics~\cite{G-Ballestero2021rev, millen2020quantum}. 

Arrays of levitated mesoscopic particles open new possibilities in both applied and fundamental physics. Arrays can enable precision force~\cite{carney2021mechanical} and force-gradient~\cite{Rudolph2022} sensing  for, e.g.,~dark matter searches~\cite{moore2021searching} or navigation~\cite{Vaccaro2017, Xue2023}, and~allow the exploration of complex collective effects~\cite{landig2016quantum,bernien2017probing}. It is proposed to generate quantum entanglement in particle arrays~\cite{rudolph2020entangling,chauhan2022tuneable}, which could be used to probe the quantum nature of gravity~\cite{marletto2017gravitationally,bose2017spin}. It is possible to individually control the particles in the array via feedback methods~\cite{vijayan2023, Siegel2025, Ren2025}. Interactions, or~coupling, between~particles can be generated via optical scattering~\cite{arita2018optical, Rieser2022, Reisenbauer2024}, an~optical cavity~\cite{vijayan2024cavity}, or the Coulomb interaction between charged objects~\cite{slezak2019microsphere,penny2023sympathetic,bykov20233d}. Coupled oscillators can be cooled via sympathetic cooling~\cite{arita2022all, penny2023sympathetic, bykov20233d}, allowing the control of elements of the particle array without the need for a direct~interaction. 

In this work, we present a levitated {synthetic} coupled oscillator formed of a real levitated charged microsphere interacting with a virtual, or~``ghost'', oscillator {simulated} via an analogue computer. {This example of hardware-in-the-loop (HIL) architecture generates a semi-virtual bi-directional coupling,} offering real-time dynamic control over the properties of the oscillators, which would not be possible with two real objects. {Since the real particle couples to its environment, our semi-virtual system could be used to probe microscale phenomena~\cite{Collin2005,Sevick_2008,Rondin2017} as a novel sensor enhanced by the tunability of the ghost particle.}

\section{Materials and Methods}
\begin{figure}
\vspace{-0.4cm}
\begin{overpic}[width=0.85\linewidth]{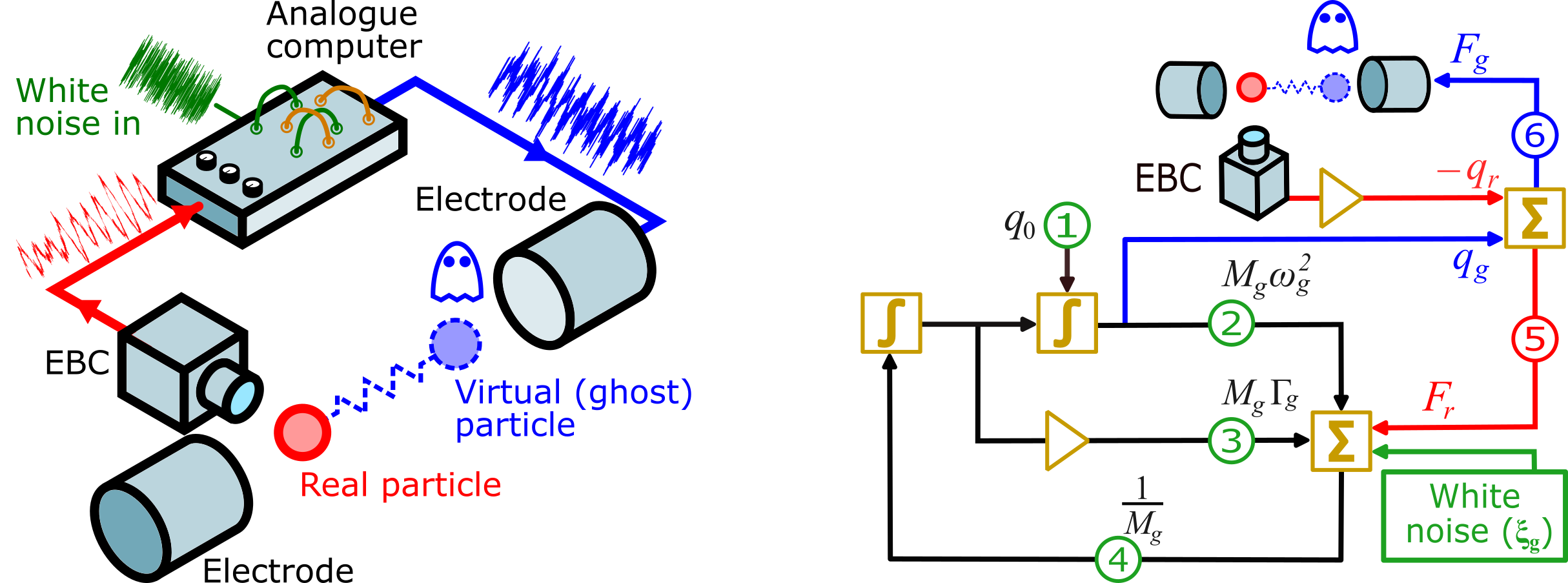}
    \put(1,35.5){(\textbf{a})}   
    \put(55,35.5){(\textbf{b})}  
\end{overpic}
\caption{(\textbf{a}) Illustration of the experiment. A~real charged microsphere is levitated in a vacuum and tracked with an event-based camera (EBC) \cite{ren2022}, producing a signal proportional to the real particle position, $q_\mathrm{r}$, which is fed into an analogue computer. The~motion of the ghost particle $q_\mathrm{g}$ is simulated on the same analogue computer and~used to produce a voltage on an electrode near the real particle. (\textbf{b}) Simplified diagram of the analogue computer and the coupling between the real and ghost particles. The~analogue computer enables dynamic control over six quantities related to the ghost particle: \circled{1} the initial position $q_0$, \circled{2} a quantity related to the oscillation frequency $\omega_\mathrm{g}$, \circled{3} a quantity related to the damping rate $\Gamma_\mathrm{g}$, \circled{4} a quantity related to the mass of the oscillator $M_\mathrm{g}$, \circled{5} the strength of the feedback force $F_\mathrm{r}$ from the real to the ghost particle, \circled{6} the strength of the feedback force $F_\mathrm{g}$ from the ghost to the real~particle. \label{fig:illustration_of_setup}}
\end{figure}   

The experimental system used to create semi-virtual coupled oscillators is illustrated in Figure \ref{fig:illustration_of_setup}{a}, where a single levitated particle (real) is coupled to a virtual particle (ghost) simulated on an analogue computer. The~real particle is a silica sphere of diameter $5\,\unit{\um}$ (Bangs Laboratories, Inc., SSD5003, Fishers, IN, USA) and positive charge of $(6 \pm 1)\times 10^3$\,e, electrically levitated in a linear Paul trap inside a vacuum chamber at a pressure of \mbox{$(2.0\pm0.4) \times 10^{-2}\,$mbar}. The~linear Paul trap has two co-axial endcap electrodes, one of which is used to apply forces to the real particle. The~centre-of-mass motion of the particle is tracked using an event-based camera (EBC) ({Prophesee, EVK1-Gen3.1 VGA, Paris, France} (camera sensor: Prophesee, PPS3MVCD, 640 $\times$ 480 pixels)), which is a neuromorphic imaging device capable of tracking multiple objects within a wide field of view with high spatial and temporal resolution. The~EBC uses an algorithm to track the real particle position in 2D, which is streamed as an analogue voltage in real time via an FPGA ({Red Pitaya, STEMlab 125-14, Solkan, Slovenia}). The details of the Paul trap geometry and EBC tracking can be found in Refs.~\cite{ren2022, Ren2025}. 

The motion of the real particle in 1D is described by the Langevin equation~\cite{Millen2020rev}:
\begin{equation}\label{equ:real_langevin_dynamics}
    \ddot{q}_\mathrm{r} = -\Gamma_\mathrm{r} \dot{q}_\mathrm{r} - \omega_\mathrm{r}^2 q_\mathrm{r} + \frac{\xi_\mathrm{r}}{M_\mathrm{r}},
\end{equation}
where $q_\mathrm{r}$ is the position of the real particle, $\Gamma_\mathrm{r}$ is its momentum damping rate, $\omega_\mathrm{r}$ is the oscillation frequency, $M_\mathrm{r}$ is its mass, and~$\xi_\mathrm{r}= \sqrt{2\Gamma_\mathrm{r} k_BT_\mathrm{r}M_\mathrm{r}}\eta(t)$ is Brownian noise due to collisions with gas molecules and residual voltage noise such that $\braket{\eta(t)}=0$ and $\braket{\eta(t)\eta(t')} = \delta(t-t')$, and~$T_\mathrm{r}$ is the centre-of-mass temperature of the real particle{, which may be above the ambient temperature due to noise in the Paul trap electronics}. We only consider motion along the direction perpendicular to the surface of the endcap electrodes. The~power spectral density (PSD) of the real particle motion has the form:
\begin{equation}\label{equ:psd_formula_real}
    S_{qq}^{\mathrm{r}}(\omega) = \frac{k_BT_\mathrm{r}\Gamma_\mathrm{r}/\pi M_\mathrm{r}}{(\omega_\mathrm{r}^2 - \omega^2)^2 + (\omega \Gamma_\mathrm{r})^2} + c_\mathrm{r},
\end{equation}
where $c_\mathrm{r}$ represents measurement noise when we record the motion $q_\mathrm{r}$ of the real~particle. 

The ghost particle's 1D dynamics are simulated on an analogue computer ({Anabrid, \textit{The Analogue Thing}, anabrid GmbH, Berlin, Germany}), which is a collection of analogue integrators, summers, multipliers, and potentiometers that can be used to continuously solve differential equations. The~analogue computer solves the Langevin equation of motion of a damped-driven harmonic oscillator in real time to simulate the ghost particle:
\begin{equation}\label{equ:single_langevin_dynamics}
    \ddot{q}_\mathrm{g} = \frac{1}{M_\mathrm{g}}\times\left(- M_\mathrm{g}\Gamma_\mathrm{g} \dot{q}_\mathrm{g} - M_\mathrm{g}\omega_\mathrm{g}^2 q_\mathrm{g} + \xi_\mathrm{g}\right),
\end{equation}
where $q_\mathrm{g}$ is the position of the ghost particle, $\Gamma_\mathrm{g}$ is the momentum damping rate, $\omega_\mathrm{g}$ is the oscillation frequency, and $M_\mathrm{g}$ is the effective mass of the oscillator. The noise term $\xi_\mathrm{g}= \sqrt{2\Gamma_\mathrm{g} k_BT_\mathrm{g}M_\mathrm{g}}\chi(t)$ is white noise added to the analogue computer to simulate Brownian motion such that $\braket{\chi(t)}=0$ and $\braket{\chi(t)\chi(t')} = \delta(t-t')$, and~$T_\mathrm{g}$ is the effective temperature of the ghost particle. {The analogue computer thus synthesises the dynamics of a damped-driven harmonic oscillator.} An illustration of the operation of the analogue computer is shown in Figure \ref{fig:illustration_of_setup}{b}, with~the detailed configuration given in Appendix \ref{apd:sec:analogue_computer}. We are able to tune the ghost particle's dynamics using potentiometers on the analogue computer: \circled{1} the initial position $q_0$, \circled{2} $M_\mathrm{g} \omega_\mathrm{g}^2$, \circled{3} $M_\mathrm{g}\Gamma_\mathrm{g}$, and \circled{4} $1/M_\mathrm{g}$. The~amplitude of the white noise $\xi_\mathrm{g}$ is controlled using a separate function generator ({Stanford Research Systems, DS335, CA, USA}).

We demonstrate a semi-virtual coupling between the real and ghost particles to simulate the dynamics of coupled harmonic oscillators. The~position signals from the real $q_\mathrm{r}$ and ghost $q_\mathrm{g}$ particles are combined and used to synthesise forces acting on each system, as~shown in Figure \ref{fig:illustration_of_setup}{b}. 
We approximate a simulation of two charged particles interacting via the Coulomb force by considering a coupling that is linearly dependent on the distance between the two particles, $F_i = k_i(q_\mathrm{r} - q_\mathrm{g})$, where $i=\{\mathrm{r},\mathrm{g}\}$ and $k_i$ is the coupling strength. {This linearised approximation is reasonable when each particle's oscillation amplitude $d$ is much smaller than their separation $\Delta R$, $d/\Delta R \ll 1$. Given that the oscillation amplitude of microparticles levitated in our Paul trap is less than $1\,\unit{\um}$, this model is valid for the physical scenario where the separation is $\gtrsim 100\,\unit{\um}$.} 

The force acting on the real particle is created by amplifying ({TTI Inc., WA301, Maisach-Gernlinden, Germany}) the voltage output $F_\mathrm{g}$ and applying it to an electrode placed $1.15\,$mm from the real particle. The~force acting on the ghost particle is implemented on the analogue computer by inverting $q_\mathrm{r}$ and summing it with $q_\mathrm{g}$, as~illustrated in Figure~\ref{fig:illustration_of_setup}{b}. Therefore, the~dynamics of the coupled semi-virtual system can be described by the equations of motion:
\begin{subequations}\label{equ:coupled_dynamics}
\begin{align}
    \ddot{q}_\mathrm{r} &= - \Gamma_\mathrm{r} \dot{q}_\mathrm{r} - \omega_\mathrm{r}^2 q_\mathrm{r} + \frac{\xi_\mathrm{r}}{M_\mathrm{r}} + \frac{k_\mathrm{r}}{M_\mathrm{r}}(q_\mathrm{g}(t-\tau) - q_\mathrm{r}),\\
    \ddot{q}_\mathrm{g} &= - \Gamma_\mathrm{g} \dot{q}_\mathrm{g} - \omega_\mathrm{g}^2 q_\mathrm{g} + \frac{\xi_\mathrm{g}}{M_\mathrm{g}} + \frac{k_\mathrm{g}}{M_\mathrm{g}}(q_\mathrm{r}(t-\tau) - q_\mathrm{g}),
  \end{align}
\end{subequations}
where $\tau$ is the {relative} time delay due to signal processing, which introduces a phase delay between the real and ghost particles' motions. 
The phase delay $\tau$ is defined relative to the particle oscillation period $T = 2\pi/\omega_\mathrm{r}$. There is a signal processing delay of less than one period~\cite{Ren2025}, which has a negligible effect on the particle dynamics due to the oscillatory nature of the motion~\cite{Debiossac2020}. This semi-virtual system simulates two harmonic oscillators coupled through a spring-like interaction. However, unlike two real particles coupled through a physical interaction, one of the oscillators is virtual and the coupling is mediated through~measurement. 

{We note that the coupled real and ghost particles' motion is only thermal in the ideal scenario of $k_{\mathrm{r}} = k_{\mathrm{g}}$ and $\tau = 0$. Equation~\eqref{equ:coupled_dynamics} does not include the measurement noise $c_r$ or electronic noise from the analogue computer and feedback loop, all of which are amplified by the couplings $k_{\mathrm{r,g}}$. The~dominant source of noise in this experiment is $c_r$ \cite{Ren2025}. We reserve a more detailed analysis of the noise in our system for a future study.}

\section{Results}
We first characterise the dynamics of the ghost oscillator simulated on the analogue computer, followed by the introduction of a coupling between the real and ghost particles. Finally, we show that, by tuning the frequency of the ghost particle, one can synthesise a tuneable coupling between the two~oscillators.

\subsection{Characterisation of the Ghost~Particle}\label{sec3.1}
We synthesise the dynamics of the ghost particle following the model in \cref{equ:single_langevin_dynamics} by varying the parameters of the analogue computer shown in Figure~\ref{fig:illustration_of_setup}{b}. We set the initial position of the ghost particle to be zero, $q_0=0$, and~record the signal proportional to the position of the ghost particle $q_\mathrm{g}(t)$.  We fit the corresponding PSD of the position \mbox{signal with}
\begin{equation}\label{equ:psd_formula_ghost}
    S_{qq}^{\mathrm{g}}(\omega) = \frac{A_\mathrm{g} \Gamma_\mathrm{g}}{(\omega_\mathrm{g}^2 - \omega^2)^2 + (\omega \Gamma_\mathrm{g})^2} + c_\mathrm{g},
\end{equation}
where $c_\mathrm{g}$ is electronic noise from the analogue computer with units of $\mathrm{V}^2/\mathrm{Hz}$. The~term $A_\mathrm{g} = \beta_\mathrm{g}^2 \times k_BT_\mathrm{g}/\pi M_\mathrm{g}$, where $\beta_\mathrm{g}$ represents the conversion between the simulated position and the output voltage of the analogue computer with units of m/V. {The temperature and the mass cannot be independently extracted from an analysis of the PSD, and~so, for this work, we leave the ghost particle signal uncalibrated.}

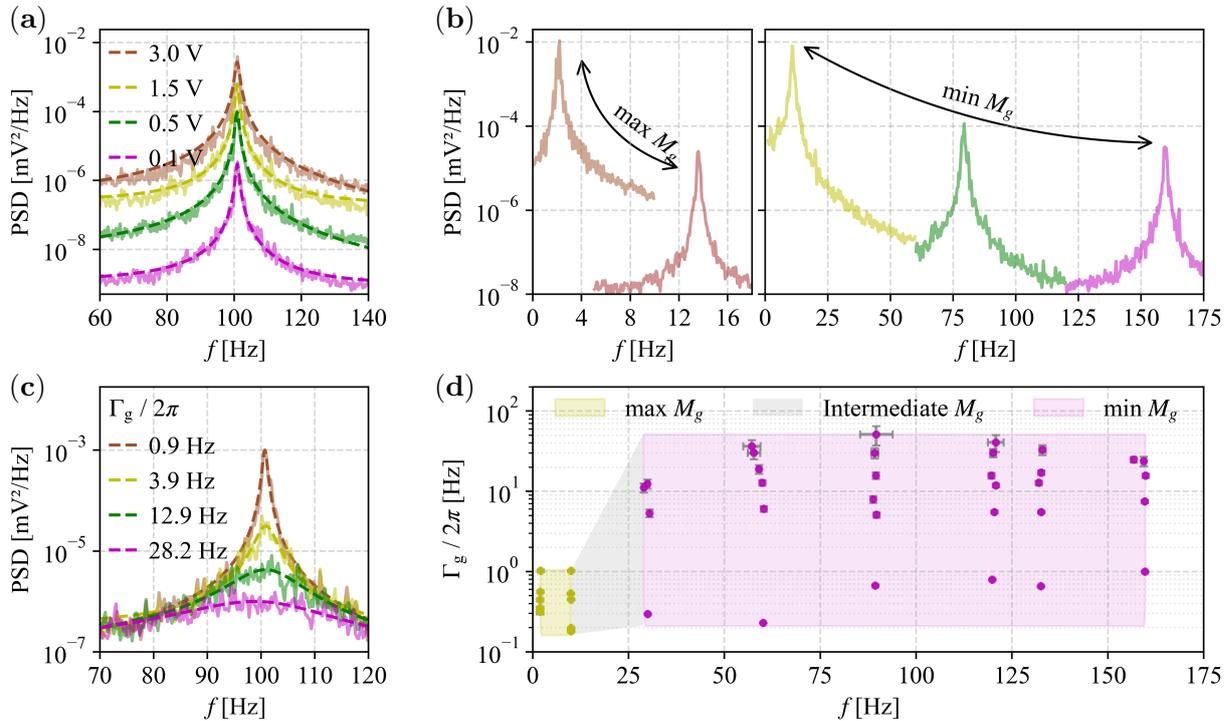
\begin{figure}[tb]
\vspace{-0.4cm}
\begin{overpic}[width=\linewidth]{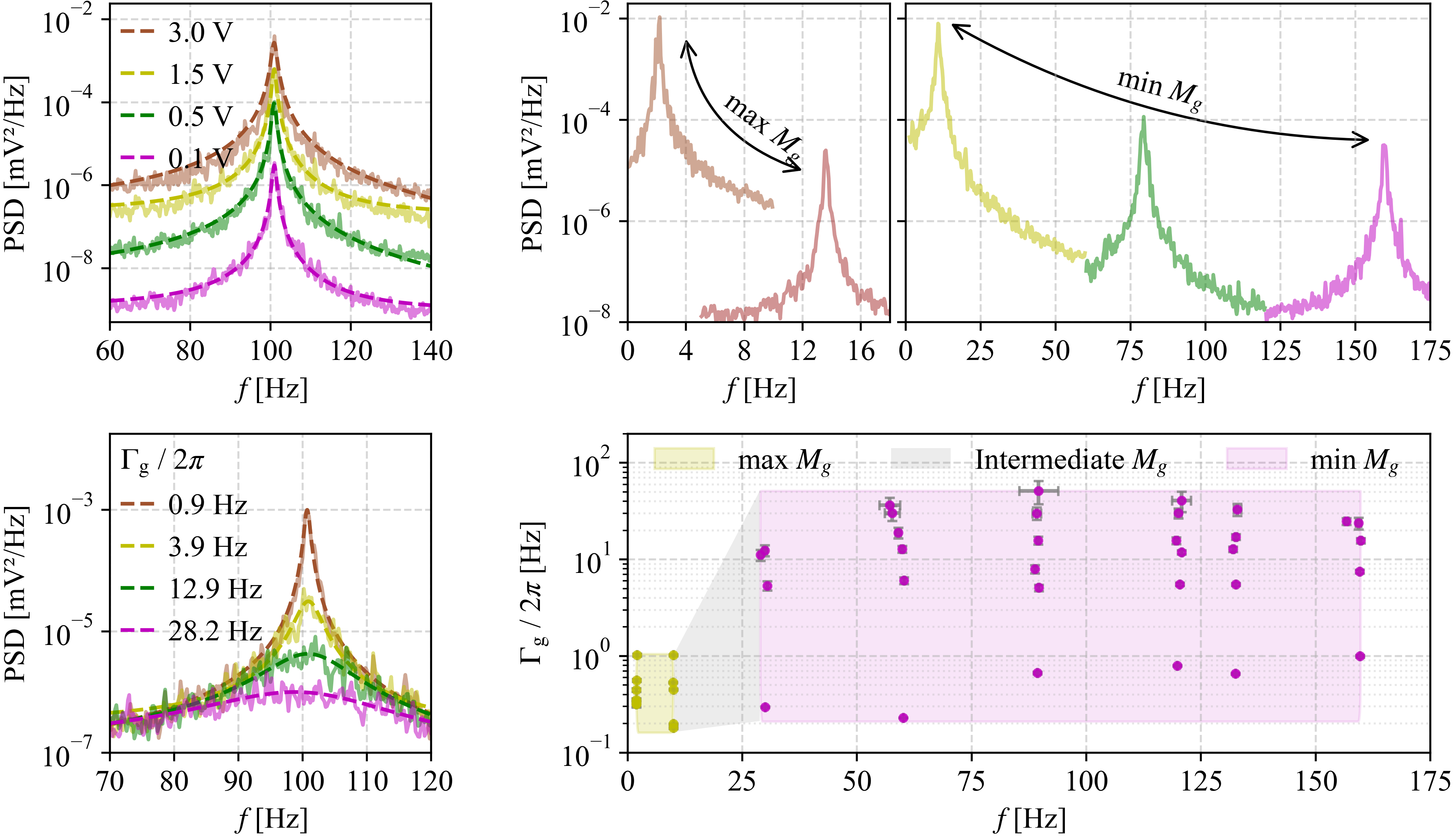}
    \put(0,57.5){(\textbf{a})}   
    \put(35,57.5){(\textbf{b})}  
    \put(0,27){(\textbf{c})}   
    \put(35,27){(\textbf{d})}  
\end{overpic}
\caption{Characterisation of the ``ghost'' virtual oscillator. (\textbf{a}) PSDs of the ghost particle's motion with a different noise amplitude $\xi_\mathrm{g}$, effectively raising the ghost particle temperature $T_\mathrm{g}$. (\textbf{b}) PSDs illustrating variable oscillation frequencies for the ghost particle. Varying the effective mass $M_\mathrm{g}$ enables different ranges of accessible frequencies. (\textbf{c}) PSDs of the ghost particle's motion with a different momentum damping rates $\Gamma_\mathrm{g}$. As~described in the text, increasing $\Gamma_\mathrm{g}$ for fixed amplitude $\xi_\mathrm{g}$ effectively reduces the temperate $T_\mathrm{g}$. (\textbf{d}) Accessible oscillation frequency and damping rate range for the ghost particle, indicated by shaded regions, with~the different colours representing different values of $M_\mathrm{g}$. Solid points represent exemplary measured values and uncertainties obtained by fitting \cref{equ:psd_formula_ghost} to records of $q_\mathrm{g}(t)$ (length $150\,\unit{s}$, sampling rate $20\,\unit{kHz}$). \label{fig:ghost_characterization}}
\end{figure}

We first illustrate the effect of changing the amplitude of $\xi_\mathrm{g}$ by changing the voltage amplitude of the input white-noise signal from $0.1\,\unit{\V}$ to $3.0\,\unit{\V}$ in Figure~\ref{fig:ghost_characterization}{a}. This represents a complex change to the ghost particle parameters, since (as seen in \cref{equ:single_langevin_dynamics}) this amplitude contains the ghost particle's mass, temperature, and damping rate. If~the other parameters remained fixed, varying the amplitude of $\xi_\mathrm{g}$ is equivalent to a change in $T_{\mathrm{g}}$. We find that the signal-to-noise ratio of the ghost oscillator on resonance is consistently around $\text{SNR}\approx30\,\unit{dB}$, and~the minimum noise floor is fixed at approximately $c_\mathrm{g} = 10^{-9}\,\unit{\mV^2/\Hz}$. 

We show the accessible resonant frequency range of the ghost system in Figure~\ref{fig:ghost_characterization}{b}, which is mass-dependent (since \circled{2} controls the term $M_\mathrm{g}\omega_\mathrm{g}$). For~the maximum accessible $M_\mathrm{g}$ the accessible frequency ranges at $\omega_\mathrm{g} = 2\pi\times (2\to 13)\,\unit{\Hz}$. For~the minimum accessible $M_\mathrm{g}$, the accessible frequency ranges at $\omega_\mathrm{g} = 2\pi\times(10\to160)\,\unit{Hz}$. We note that the noise floor increases at low frequencies due to $1/f$ noise in the analogue~computer.

Next, we control the momentum damping rate of the ghost system from $\Gamma_\mathrm{g} = 2\pi \times (0.8 \to 27.5)\,\unit{Hz}$ (using the control \circled{3} to adjust the term $M_\mathrm{g}\Gamma_\mathrm{g}$ ), as~shown in Figure~\ref{fig:ghost_characterization}{c}. The~lower limit of the damping rate is comparable to that of our real particle, which is approximately $(2\pi\times1)\,\unit{Hz}$ at $2\times10^{-2}\,\unit{mbar}$. As~we increase $\Gamma_\mathrm{g}$, the~peak of the PSD decreases accordingly until it reaches the noise floor. By~increasing the parameter $M_\mathrm{g} \Gamma_\mathrm{g}$ while keeping the amplitude of the noise $\xi_\mathrm{g}$ constant, the~ghost system's effective temperature decreases since $k_b T_\mathrm{g} = \xi_\mathrm{g}^2/(4M_\mathrm{g}\Gamma_\mathrm{g})$, as~seen by the decrease in area under the PSD in Figure~\ref{fig:ghost_characterization}{c}.

Finally, we summarise the accessible oscillation frequencies and momentum damping rate of our ghost system in Figure~\ref{fig:ghost_characterization}{d}. The~accessible parameters depend on the effective mass of the ghost particle $M_\mathrm{g}$. At~the minimum value of $M_\mathrm{g}$, the parameter range is broad and overlaps with the parameters of our real particle. We will show that this enables us to engineer a semi-virtual coupling between the real and ghost oscillators. The~high degree of controllability over the ghost particle properties will enable the generation of coupled oscillators, which would be difficult to engineer in a purely real two-particle~system.

\subsection{Coupling the real and ghost particles}
\begin{figure}[tb]
\vspace{-0.4cm}
\begin{overpic}[width=\linewidth]{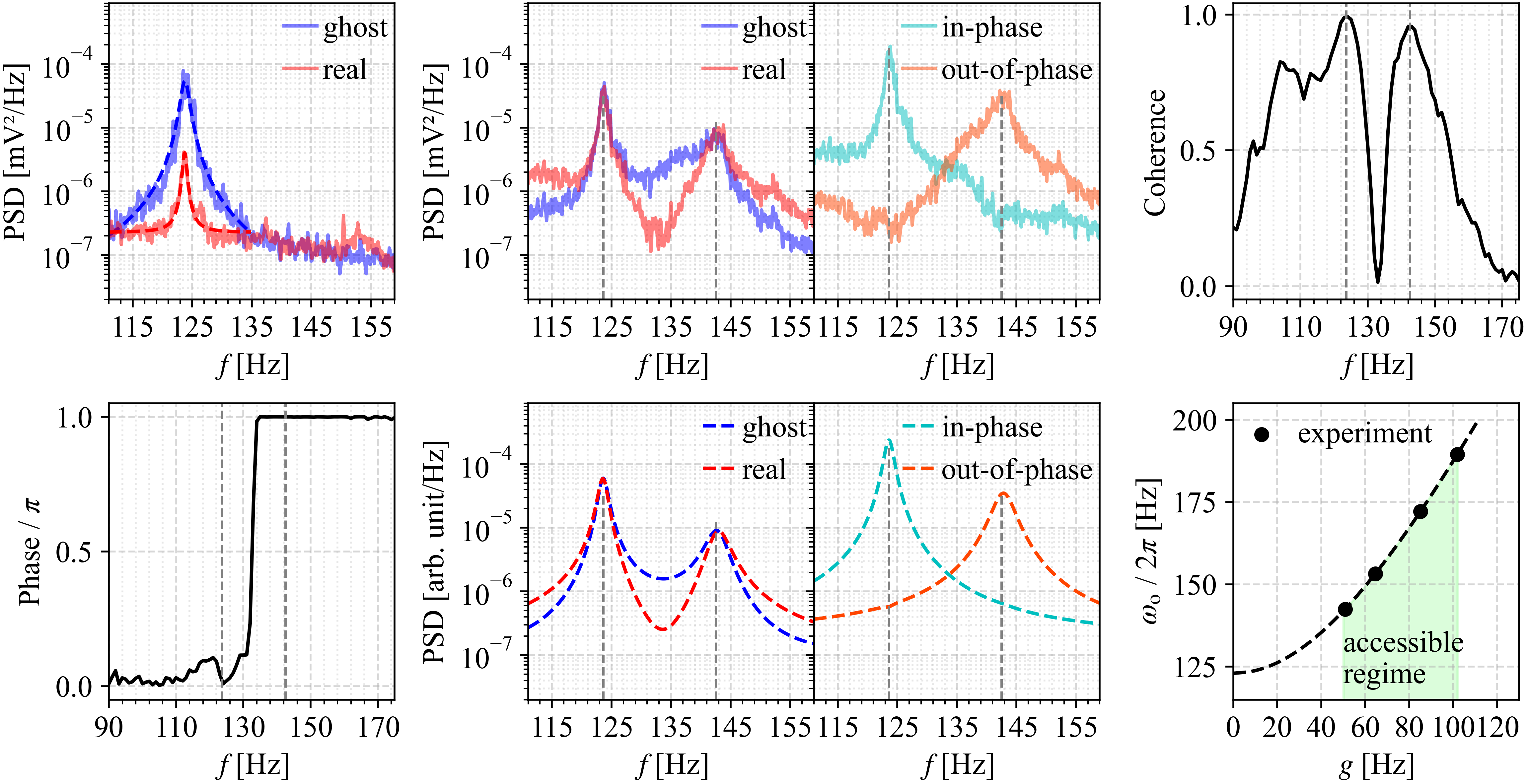}
    \put(0,50){(\textbf{a})} 
    \put(27.5,50){(\textbf{b})} 
    \put(74,50){(\textbf{c})} 
    \put(0,24){(\textbf{d})} 
    \put(27.5,24){(\textbf{e})} 
    \put(74,24){(\textbf{f})} 
\end{overpic}
\caption{Synthesising
 a coupling between the real and ghost particles. (\textbf{a}) Position PSDs of the uncoupled real (red) and ghost (blue) particles, $\omega_\mathrm{r} = 2\pi\times (123\pm0.2)\,$Hz, $\omega_\mathrm{g} = 2\pi\times (123\pm0.2)\,$Hz, $\Gamma_\mathrm{r} = 2\pi\times (1.0\pm0.2)\,$Hz, $\Gamma_\mathrm{g} = 2\pi\times (1.2\pm0.2)\,$Hz. These parameters and their uncertainties are extracted by fitting \cref{equ:psd_formula_real,equ:psd_formula_ghost} to the data. (\textbf{b}) (Left) After the semi-virtual oscillators are coupled, both PSDs exhibit similar behaviour with resonances at 123.0\,Hz and 142.5\,Hz. (Right)  PSDs of the in-phase (cyan) and out-of-phase (orange) motion of the coupled semi-virtual particles, identifying two distinct collective modes. {On subfigures (b-e) we indicate 123.0 Hz and 142.5 Hz with a vertical dashed line.} (\textbf{c}) Coherence and (\textbf{d}) phase between the real and ghost particle's motion, illustrating that the modes are coupled, and~the nature of each mode, respectively. (\textbf{e}) Simulation of the data in (\textbf{b}), showing good quantitative agreement. (\textbf{f}) Relationship between the coupling strength $g$ and {the out-of-phase frequency $\omega_\mathrm{o}$}. The~shaded region illustrates the values of $g$ accessible in our system, and~the solid points indicate the values we have measured in our~experiment. \label{fig:symmetric_coupling}}
\end{figure}
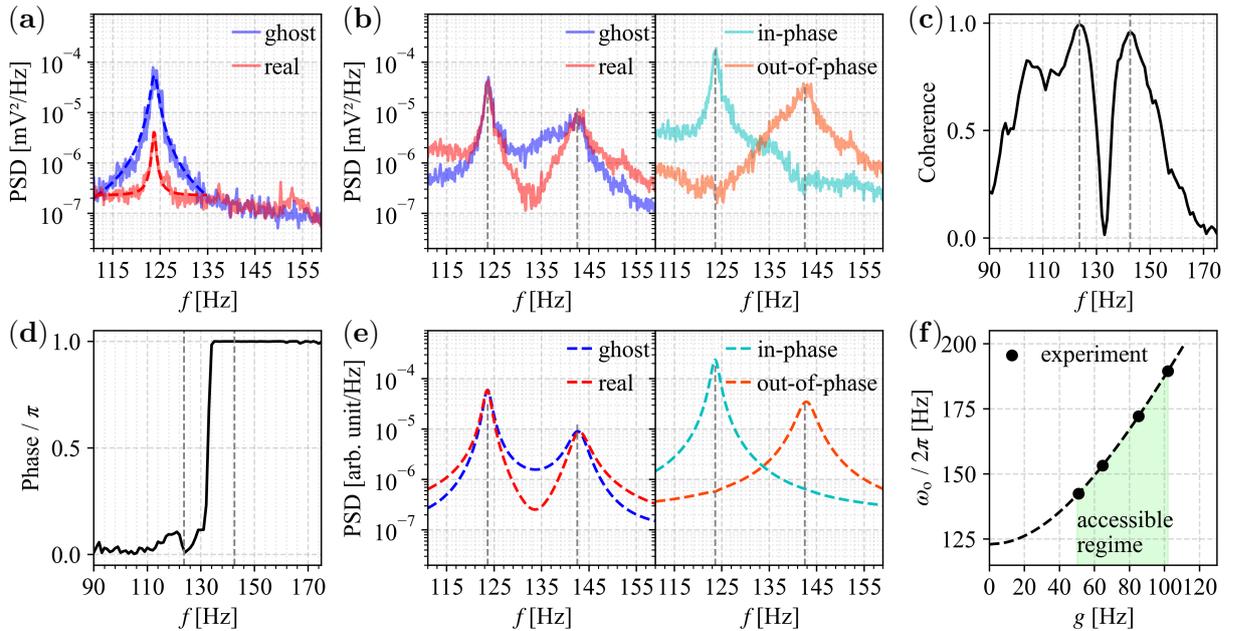 

Once the real and ghost particles are coupled, the~PSDs for the coupled system described by \cref{equ:coupled_dynamics} are given by the matrices:
\begin{equation}
    \mathbf{S}_{qq} =
    \begin{pmatrix}
        S_{qq}^\mathrm{r,r}(\omega) & S_{qq}^\mathrm{r,g}(\omega) \\
        S_{qq}^\mathrm{g,r}(\omega) & S_{qq}^\mathrm{g,g}(\omega)
    \end{pmatrix},
    \;
    \mathbf{S}'_{qq} =\mathbf{T}\mathbf{S}_{qq}\mathbf{T}=
    \begin{pmatrix}
        S_{qq}^\mathrm{i,i}(\omega) & S_{qq}^\mathrm{i,o}(\omega) \\
        S_{qq}^\mathrm{o,i}(\omega) & S_{qq}^\mathrm{o,o}(\omega)
    \end{pmatrix},
    \;
    \text{with }
    \mathbf{T} =
    \begin{pmatrix}
        1 & 1 \\
        1 & -1
    \end{pmatrix},
\end{equation}
where the first term relates to the real and ghost particles' positions $\{q_\mathrm{r},\ q_\mathrm{g}\}$ and the second term to the in-phase and out-of-phase modes $\{q_\mathrm{i}=q_\mathrm{r}+q_\mathrm{g},\ q_\mathrm{o}=q_\mathrm{r}-q_\mathrm{g}\}$.
The $\{q_\mathrm{i},\ q_\mathrm{o}\}$ modes are normal modes for the system if two coupled harmonic oscillators are identical ($S_{qq}^\mathrm{i,o}$ = 0, e.g.,~refs.~\cite{penny2023sympathetic,bykov20233d}). We have a high degree of control over the ghost particle properties, such that we can set $\Gamma_\mathrm{g} = \Gamma_\mathrm{r}$ and $\omega_\mathrm{g}=\omega_\mathrm{r}$ and decouple $\{q_\mathrm{i}, q_\mathrm{o}\}$ by tuning the coupling strengths $k_\mathrm{r}$ and $k_\mathrm{g}$. 

In Figure~\ref{fig:symmetric_coupling}{a}, we show the position PSDs for uncoupled real and ghost particles $\{S_{qq}^\mathrm{r,r}(\omega),$ $S_{qq}^\mathrm{g,g}(\omega)\}$, with similar resonant frequencies at $\omega_\mathrm{r}\approx\omega_\mathrm{g}\approx2\pi\times 123\,\unit{\Hz}$ and damping $\Gamma_\mathrm{r}\approx\Gamma_\mathrm{g}\approx2\pi\times 1\,\unit{\Hz}$. Introducing the coupling as described in \cref{equ:coupled_dynamics} changes the corresponding position PSDs to the one shown in Figure~\ref{fig:symmetric_coupling}{b} (left) , with~the appearance of a mode at $\sim142.5$\,Hz. We calculate the in-phase and out-of-phase modes $\{q_\mathrm{i},\ q_\mathrm{o}\}$ and show their PSDs in Figure~\ref{fig:symmetric_coupling}{b} (right). We adjust $k_\mathrm{r}$ and $k_\mathrm{g}$ until $S_{qq}^\mathrm{i,i}(\omega)$ only shows a peak at 123.0\,Hz, and~$S_{qq}^\mathrm{o,o}(\omega)$ only shows a peak at 142.5\,Hz, corresponding to a reciprocal coupling. 

To further explore the nature of the interaction, we calculate the coherence and relative phase of the coupled real and ghost particles' motion $\{q_\mathrm{r},\ q_\mathrm{g}\}$, defined as
\begin{equation}
        \text{Coherence} = \frac{\abs{S_{qq}^\mathrm{r,g}(\omega)}^2}{S_{qq}^\mathrm{r,r}(\omega)S_{qq}^\mathrm{g,g}(\omega)} 
        \qquad \text{and} \qquad
        \text{Phase} = \mathrm{arg}[S_{qq}^\mathrm{r,g}(\omega)].
\end{equation}
Here $\mathrm{arg}$ is the argument function that measures the angle of a complex number to the positive real axis. The~calculated coherence and relative phase are shown in \mbox{Figure~\ref{fig:symmetric_coupling}{c} and \ref{fig:symmetric_coupling}{d},} respectively. The~coherence approaches unity at the frequencies 123.0\,Hz and 142.5\,Hz, indicating that these two modes are coupled. There is a sharp change by $\pi$ radians in the relative phase of oscillations across the same range, indicating that they are indeed in-phase and out-of-phase~modes.

We verify our conclusions regarding the dynamics by comparing the data to the theoretical model derived from \cref{equ:coupled_dynamics} and given in Appendix \ref{apd:sec:PSDs_for_coupled_systems}. The~theoretical PSDs are shown in Figure~\ref{fig:symmetric_coupling}{d} for $\{S_{qq}^\mathrm{r,r}(\omega),\ S_{qq}^\mathrm{g,g}(\omega)\}$ (left) and $\{S_{qq}^\mathrm{i,i}(\omega),\ S_{qq}^\mathrm{o,o}(\omega)\}$ (right). We find $\omega_\mathrm{r,g}$ and $\Gamma_\mathrm{r,g}$ by fitting the experimental data when the particles are uncoupled. The~model correctly predicts the normal-mode frequencies at 123.0\,Hz and 142.5\,Hz. The~size of the dip in-between the modes is due to the noise $\xi_\mathrm{r}/M_\mathrm{r},\ \xi_\mathrm{g}/M_\mathrm{g}$. These noise terms are not equal, leading to different sized dips, as~reproduced by the model. Furthermore, the~model tells us that the coupling strengths $k_\mathrm{r}/M_\mathrm{r}$ and $k_\mathrm{g}/M_\mathrm{g}$ are approximately equal, evidenced by the similar peak heights in the PSDs of each particle. The~difference between the peak heights of the in-phase and out-of-phase modes is due to a measured time delay of $\tau=1.1\,\unit{\ms}$ in our electronics, also reproduced by the model. With~this, we are confident that the dynamics we see truly reflect those of coupled oscillators, and~our model well describes what is seen in the~experiment.

We estimate the quadratic mean coupling rate $g$ of our coupled system, which we define as follows:
\begin{equation}\label{equ:coupling_rate_fomula}
    g = \sqrt{\frac{1}{2}\left(\omega_\mathrm{o}^2-\omega_\mathrm{i}^2\right)},
\end{equation}
where $\omega_\mathrm{i,o}$ are the frequencies of in-phase and out-of-phase modes. This coupling rate $g$ only depends on the coupling strengths $k_\mathrm{r}/M_\mathrm{r}$ and $k_\mathrm{g}/M_\mathrm{g}$ for our linearly coupled system in the weak-damping regime (see Appendix~\ref{apd:sec:quadratic_mean_of_coupling}). In~Figure~\ref{fig:symmetric_coupling}{f}, we show experimental data where we keep the in-phase mode $\omega_\mathrm{i}=2\pi\times123\,\unit{\Hz}$ and change the out-of-phase mode frequency $\omega_\mathrm{o}$ by tuning $k_\mathrm{r}$ and $k_\mathrm{g}$ with fixed $M_\mathrm{g}$. In~this scenario, the coupling rate $g$ ranges from $2\pi\times50\,\unit{\Hz}$ {(this minimum coupling rate corresponds to the minimum output of our analogue computer that is non-zero)} to $2\pi\times102\,\unit{\Hz}$.

Finally, we make a note about the calibration of the semi-virtual coupled oscillators. While it is straightforward to calibrate the real particle motion~\cite{ren2022} and~feasible to calibrate the uncoupled ghost particle motion (see Section~\ref{sec3.1}), calibrating the coupled ghost particle motion is ambiguous. The~interaction between the two oscillators is synthetic; hence, any notion of ghost particle charge or~real--ghost particle separation is abstract.

\subsection{Tunable coupling with different frequency oscillators}

In our semi-virtual coupled-oscillator system, we have the ability to widely and dynamically vary the properties of the ghost particle, in~a manner and over a range that would be challenging to achieve with two real particles in a standard levitation experiment. Here, we demonstrate a coupling between the semi-virtual oscillators even when the difference in their resonant frequencies is~large.

\begin{figure}[tb]
\vspace{-0.4cm}
\begin{overpic}[width=\linewidth]{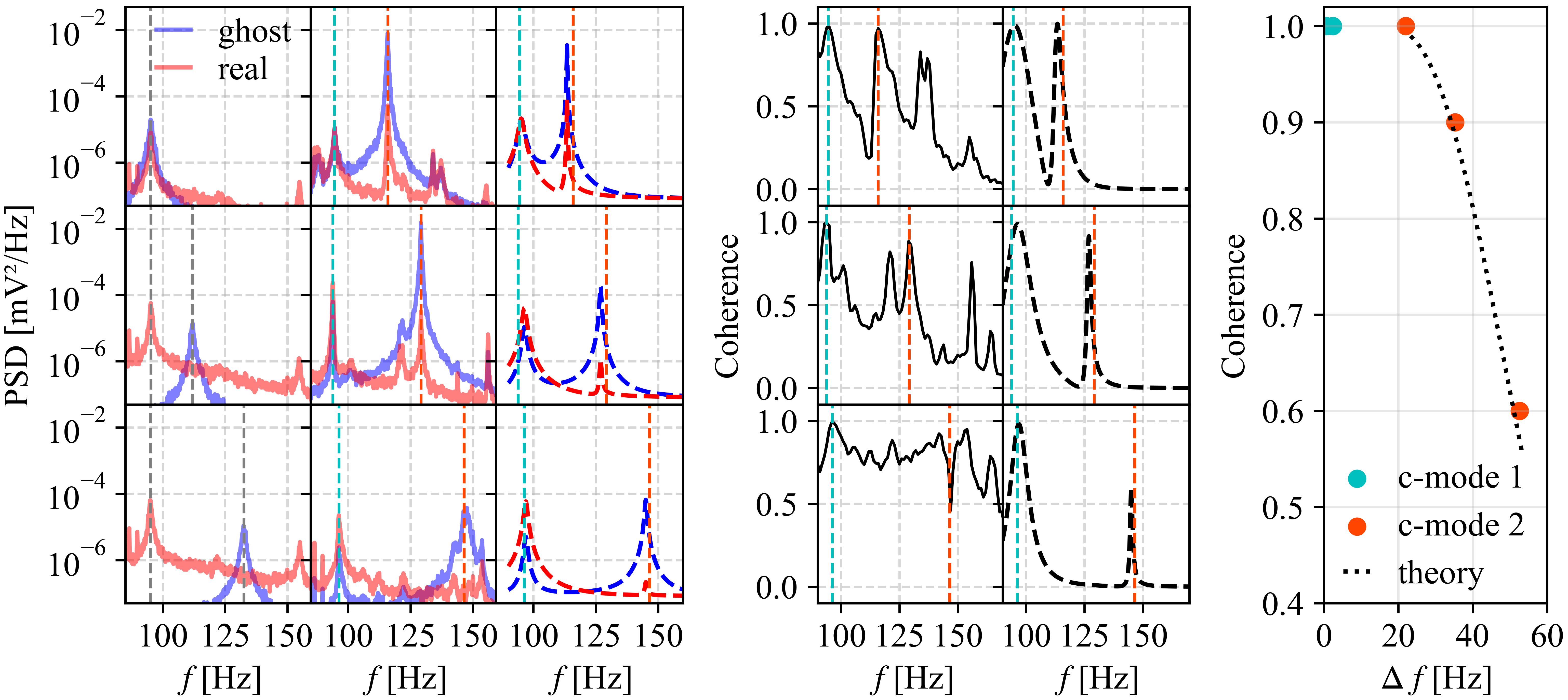}
    \put(0,42.5){(\textbf{a})} 
    \put(45,42.5){(\textbf{b})}
    \put(77.5,42.5){(\textbf{c})} 
\end{overpic}
\caption{Interaction between a real and ghost particle with different frequencies. (\textbf{a}) Position PSDs of the real particle with a fixed frequency (red) and a ghost particle  with a variable frequency (blue), when they are uncoupled (left) and coupled (middle). The~simulations (right) show good quantitative agreement. {Grey vertical dashed lines indicate the positions of the uncoupled modes. Cyan and orange vertical dashed lines indicate the positions of c-modes 1 and 2, respectively.} (\textbf{b}) Calculated (left) and simulated (right) coherence between the coupled real and ghost particle motion, with~\textit{c}-mode 1 and \textit{c}-mode 2 marked by {cyan} and orange dotted lines, respectively. (\textbf{c}) Variation in the coherence at the position of \textit{c}-mode 1 (cyan) and \textit{c}-mode 2 (orange) as a function of the mode separation $\Delta f$, compared to theory (dotted black line). \label{fig:tunable_coupling}}
\end{figure}

In Figure~\ref{fig:tunable_coupling}{a}, we show the position PSDs of the real and ghost particles when they are uncoupled (left). We fix the real particle frequency and show three cases where the ghost particle frequency varies between ($95\to132$)\,Hz. The~real particle damping rate is $\Gamma_\mathrm{r} = 2\pi\times (1.5\pm0.5)\,$Hz, and we set $\Gamma_\mathrm{g} = 2\pi\times (1.3\pm0.3)\,$Hz. 

We show the position PSDs of the real and ghost particles after the semi-virtual oscillators are coupled in Figure~\ref{fig:tunable_coupling}{a} (middle). There are two distinct coupled modes, which we call \textit{c}-modes 1 and 2, which can no longer be thought of as in-phase and out-of-phase due to the frequency difference. They appear in the PSDs of both particles and are marked by cyan and orange dotted lines. We present a comparison to our theoretical model in Figure~\ref{fig:tunable_coupling}{a} (right).  The~model correctly predicts the frequencies of the \textit{c}-modes and~suggests that the system has an asymmetric coupling $k_\mathrm{g}/M_\mathrm{g}>k_\mathrm{r}/M_\mathrm{r}$, evidenced by the unequal peak heights of the \textit{c}-modes in each~PSD.

To explore the nature of the coupling, we calculate the coherence between the coupled real and ghost motions, as shown in Figure~\ref{fig:tunable_coupling}{b} (left). For~this scenario, the~coupled signals are less visible above the noise than in the case of equal-frequency oscillators, leading to a less clean coherence spectrum. We note that the visibility of \textit{c}-mode~2 decreases with increasing frequency mismatch $\Delta f$. The~corresponding theoretical coherence spectrum is shown in Figure~\ref{fig:tunable_coupling}{b} (right), which reproduces the drop in coherence of \textit{c}-mode~2. In~the model, this loss of coherence is caused by the relative time delay $\tau=1.1~\unit{\ms}$ in the system, which shifts the theoretical resonant frequencies to the complex plane (see Appendix~\ref{apd:sec:quadratic_mean_of_coupling}). We consider the coherence only at the frequencies of the \textit{c}-modes in Figure~\ref{fig:tunable_coupling}{c} and~compare it to our model. We see excellent quantitative agreement when considering the loss of coherence with frequency difference, despite the noisier spectra in Figure~\ref{fig:tunable_coupling}{a,b}.

\section{Discussion}
We have demonstrated tuneable coupling between two oscillators, one of which is a real charged microsphere levitated in vacuum and the other is a near-identical ghost oscillator synthesised on an analogue computer. This proof-of-principle work represents a new toolbox for levitated particle manipulation by~considering it as hardware-in-the-loop (HIL) of an electronic control system. {In the work presented here, the~analogue computer could be replaced by a digital system, which would offer advantages in terms of flexibility and noise. Analogue computers have specific advantages, for~example, when simulating the ``fast--slow'' dynamics that are found throughout nature~\cite{BERTRAM2017105}.}

Charged particles levitated in a Paul trap are an ideal platform for physical analogue simulation, due to a high degree of environmental isolation in vacuum, and~the possibility to engineer a user-defined environment via electrical forces~\cite{Message2025}. The~real-time feedback we use to generate an interaction between a single real particle and the analogue ghost is scaleable to arrays of particles or multiple degrees-of-freedom simultaneously~\cite{Ren2025}.
In this work, we have only varied the relative oscillation frequency between the semi-virtual coupled oscillators, as well as~the magnitude of linear coupling between the particle. Future work will consider oscillators with different properties, for~example, by synthesizing enhanced dissipation for the ghost to enable sympathetic cooling of the real particle. By~unbalancing the interaction between the semi-virtual oscillators ($k_\mathrm{r} \neq k_\mathrm{g}$), it will be possible to synthesise non-reciprocal Coulomb interactions in our HIL system, which cannot be physically generated. Finally, we have only explored a linear coupling between the oscillators, yet it is possible to synthesise arbitrary interactions, which may potentially be unfeasible to digitally simulate. This work is in the tradition of physical analogue engineering, which can be used to explore inaccessible areas of physics~\cite{Svancara2024} and enable efficient information processing~\cite{Wright2022}. 

\paragraph*{Author Contributions}
R.Y. carried out experiments and analysed data. Y.R. designed and carried out experiments. D.Y.S., A.S., and J.D.P. programmed and characterised the analogue computer. Q.W. developed the physical models and analysed the data. J.M. designed the experiment and conceived of the project. R.Y., Y.R., Q.W., and J.M. wrote the~manuscript. 

\paragraph*{Funding}
This project was supported by Leverhulme Research Grant RPG-2024-017 and~by the European Research Council (ERC) under the EU's Horizon Europe programme (Grant Agreement No. 803277). We appreciate support from King's College London through the King's Undergraduate Research Fellow (KURF) scheme. 

\paragraph*{Data Availability}
Data will be available on the KORDS service upon publication.

\paragraph*{Acknowledgments}
We appreciate technical assistance from Jessamine Particle.

\newpage
\appendix
\makeatletter
\renewcommand{\@seccntformat}[1]{%
  \appendixname~\csname the#1\endcsname\quad}
\makeatother
\section{Configuration of the analogue computer}
\renewcommand{\thefigure}{A\arabic{figure}}

\label{apd:sec:analogue_computer}
\begin{figure}[tb]
\includegraphics[width=0.95\linewidth]{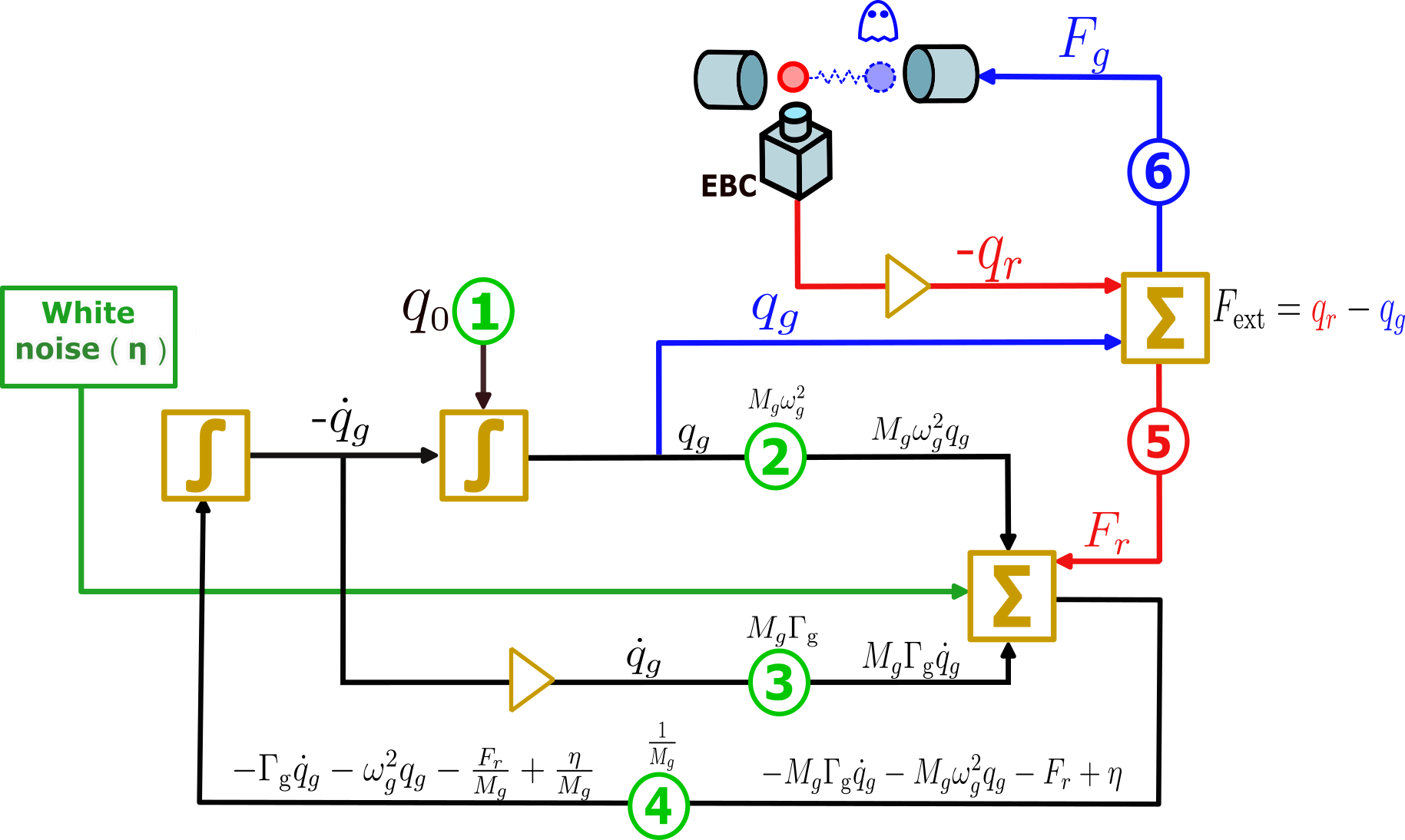}
\caption{Simulating the dynamics of a levitated microparticle on an analogue computer by solving its equations of motion. A~description is given in the~text.} \label{fig:circuit diagram}
\end{figure}

In \cref{fig:circuit diagram}, the~\scalebox{0.7}{\fbox{$\int$}} symbols denote integrators that time-integrate the sum of
their inputs with an implicit sign inversion. The~symbol
\scalebox{1.5}{$\displaystyle\triangleright$} denotes an inverter (sign flip), and~\scalebox{0.8}{\fbox{$\Sigma$}} denotes a summer that adds its inputs and implicitly
inverts the output sign. Numbers in circles represent variable potentiometers that set parameters for the ghost particle and the
forces acting on both real and ghost particles: \circled{1}~sets the initial value $q_0$; \circled{2}~sets $M\mathrm{g}\omega\mathrm{g}^2$; \circled{3}~sets $M\mathrm{g}\Gamma\mathrm{g}$; \circled{4}~sets $1/M\mathrm{g}$; and \circled{5}--\circled{6}~set the force input to the real and ghost particles, respectively.

The upper summer forms a Coulomb-like interaction from the relative displacement of the particles. Because~the analogue computer lacks the ability to divide two signals, we linearise the Coulomb force using a first-order Taylor expansion about the operating point, which is justified by the small oscillation amplitudes in the experiment. The~lower summer combines the terms of the Langevin equation; its output passes through \circled{4} to produce the acceleration $\ddot{q}$. Double integration then yields the ghost particle position signal. {The range of parameters is limited by the design of the analogue computer used in this study, and~ultimately by the precision and power handling of the components used.}

\section{PSDs for coupled harmonic oscillators}\label{apd:sec:PSDs_for_coupled_systems}

By rearranging \cref{equ:coupled_dynamics} and taking the Fourier transform, the~coupled system in the frequency domain {reads} \begin{subequations}
\begin{align}
    -\omega^2 q_\mathrm{r}(\omega) + i\omega \Gamma_\mathrm{r} q_\mathrm{r}(\omega) + \left(\omega_\mathrm{r}^2 + \frac{k_\mathrm{r}}{M_{r}}\right) q_\mathrm{r} (\omega) -  e^{-i\omega\tau}\frac{k_\mathrm{r}}{M_{r}} q_\mathrm{g}(\omega) &=\frac{\xi_\mathrm{r}(\omega)}{M_\mathrm{r}},\\
    -\omega^2 q_\mathrm{g}(\omega) + i\omega \Gamma_\mathrm{g} q_\mathrm{g}(\omega) + \left(\omega_\mathrm{g}^2 + \frac{k_\mathrm{g}}{M_{g}}\right) q_\mathrm{g} (\omega) -  e^{-i\omega\tau}\frac{k_\mathrm{g}}{M_{g}} q_\mathrm{r}(\omega) &=\frac{\xi_\mathrm{g}(\omega)}{M_\mathrm{g}}.
  \end{align}
\end{subequations}
For simplicity, this system of equations can be written in the matrix format $\mathbf{D}(\omega)\mathbf{q}(\omega) = \mathbf{M} {\bm \xi}(\omega)$ with $\mathbf{q}(\omega) = \left(q_\mathrm{r}(\omega), q_\mathrm{g}(\omega)\right)^T$,  ${\bm \xi}(\omega) = \left(\xi_\mathrm{r}(\omega), \xi_\mathrm{g}(\omega)\right)^T$ and
\begin{equation}\label{apd:equ:D_and_M_matrices}
    \mathbf{D}(\omega) = 
    \begin{pmatrix}
        -\omega^2+i\omega\Gamma_\mathrm{r} + \left(\omega_\mathrm{r}^2 + \frac{k_\mathrm{r}}{M_{r}}\right) & -e^{-i\omega\tau}\frac{k_\mathrm{r}}{M_{r}} \\
        -e^{-i\omega\tau}\frac{k_\mathrm{g}}{M_{g}} & -\omega^2 + i \omega \Gamma_\mathrm{g} +\left(\omega_\mathrm{g}^2 + \frac{k_\mathrm{g}}{M_{g}}\right)
    \end{pmatrix}, \;
    \mathbf{M} = 
    \begin{pmatrix}
        \frac{1}{M_\mathrm{r}} & 0 \\
        0 & \frac{1}{M_\mathrm{g}}
    \end{pmatrix}.
\end{equation}
This is a linear system with the solution
\begin{equation}
    \mathbf{q}(\omega) = \mathbf{G}(\omega) {\bm \xi}(\omega)
\end{equation}
where $\mathbf{G}(\omega) = \mathbf{D}(\omega)^{-1}\mathbf{M}$ and
\begin{equation}
    \mathbf{D}(\omega)^{-1} = \frac{1}{\Delta(\omega)}
    \begin{pmatrix}
        D_\mathrm{gg}(\omega) & -e^{-i\omega\tau}\frac{k_\mathrm{r}}{M_\mathrm{r}} \\ 
        -e^{-i\omega\tau}\frac{k_\mathrm{g}}{M_\mathrm{g}} & D_\mathrm{rr}(\omega)
    \end{pmatrix},
\end{equation}
with $D_\mathrm{rr}(\omega) =  -\omega^2+i\omega\Gamma_\mathrm{r} + \left(\omega_\mathrm{r}^2 + \frac{k_\mathrm{r}}{M_{r}}\right)$, $D_\mathrm{gg}(\omega) = -\omega^2 + i \omega \Gamma_\mathrm{g} +\left(\omega_\mathrm{g}^2 + \frac{k_\mathrm{g}}{M_{g}}\right)$ and $\Delta(\omega) = D_\mathrm{rr}(\omega)D_\mathrm{gg}(\omega) - e^{-2i\omega\tau}k_\mathrm{r}k_\mathrm{g}/M_\mathrm{r}M_\mathrm{g}$ that one can read from \cref{apd:equ:D_and_M_matrices}.

\begin{figure}[tb]
\vspace{-0.4cm}
\includegraphics[width=.99\linewidth]{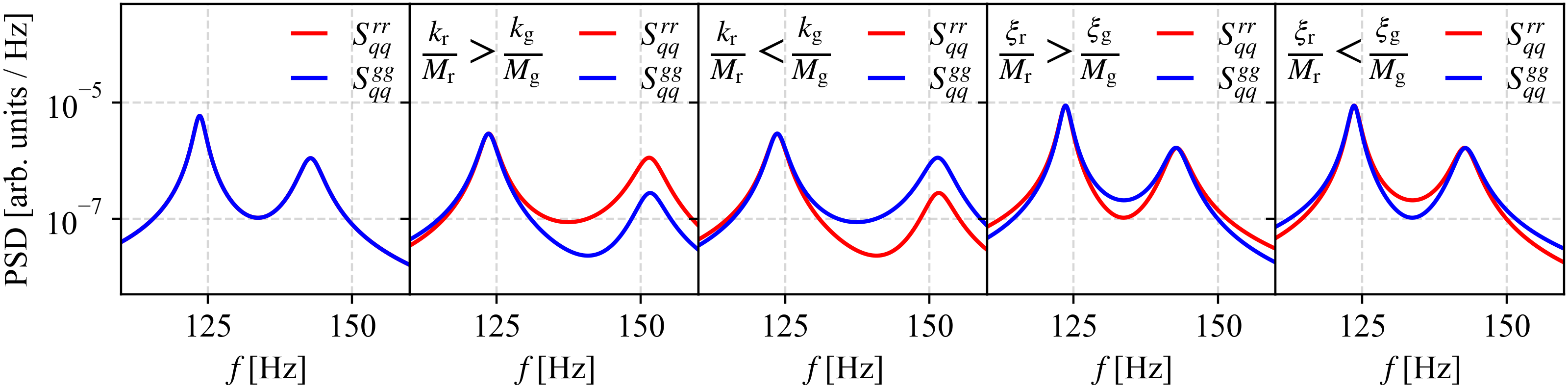}
\caption{{Illustration} of PSDs $S_{qq}^\mathrm{r,r} (\omega)$ and $S_{g}^\mathrm{r,r} (\omega)$ under asymmetric coupling $k_\mathrm{r,g}/M_\mathrm{r,g}$ and noise $\xi_\mathrm{r,g}/M_\mathrm{r,g}$. \label{apd:fig:asymetry_settings}}
\end{figure} 

The PSDs of the coupled system is given by $S_{qq}^\mathrm{i,j} (\omega) = \braket{q_\mathrm{i}(\omega) q^\ast_\mathrm{j}(\omega)}$ for $\mathrm{i}, \mathrm{j} = \mathrm{r}, \mathrm{g}$, and~it reads in matrix format as
\begin{align}
    \mathbf{S}_{qq} &= \braket{\mathbf{q}(\omega) \mathbf{q}(\omega)^\dagger}, \notag\\
    &=\mathbf{G}(\omega)\braket{{\bm \xi}(\omega){\bm \xi}(\omega)^\dagger} \mathbf{G}(\omega)^\dagger, \notag \\
    &=\mathbf{G}(\omega) \mathbf{S}_{\xi\xi} \mathbf{G}(\omega)^\dagger.
\end{align}
Given the white noise definition, $\mathbf{S}_{\xi\xi}$ is a diagonal matrix with the diagonal terms $\braket{\xi_\mathrm{r}^2}$ and $\braket{\xi_\mathrm{g}^2}$. This gives the PSDs for each mode and the cross {term,} 
\begin{subequations}
\begin{align}
        S_{qq}^\mathrm{r,r} (\omega) &= \frac{1}{\abs{\Delta(\omega)}^2}\left( \abs{D_\mathrm{gg}(\omega)}^2 \frac{\braket{\xi_\mathrm{r}^2}}{M_\mathrm{r}^2} + \frac{k_\mathrm{r}^2}{M_\mathrm{r}^2} \frac{\braket{\xi_\mathrm{g}^2}}{M_\mathrm{g}^2}\right), \\
        S_{qq}^\mathrm{g,g} (\omega) &= \frac{1}{\abs{\Delta(\omega)}^2}\left( \abs{D_\mathrm{rr}(\omega)}^2 \frac{\braket{\xi_\mathrm{g}^2}}{M_\mathrm{g}^2} + \frac{k_\mathrm{g}^2}{M_\mathrm{g}^2} \frac{\braket{\xi_\mathrm{r}^2}}{M_\mathrm{r}^2}\right), \\
        S_{qq}^\mathrm{r,g} (\omega) &= \frac{1}{\abs{\Delta(\omega)}^2} \left( e^{i\omega\tau}D_\mathrm{gg}(\omega) \frac{k_\mathrm{g}}{M_\mathrm{g}} \frac{\braket{\xi_\mathrm{r}^2}}{M_\mathrm{r}^2} + e^{-i\omega\tau}D_\mathrm{rr}^\ast(\omega) \frac{k_\mathrm{r}}{M_\mathrm{r}}\frac{\braket{\xi_\mathrm{g}^2}}{M_\mathrm{g}^2}\right), \\
        S_{qq}^\mathrm{g,r} (\omega) &= S_{qq}^\mathrm{r,g} (\omega)^\ast,
    \end{align}
\end{subequations}
which returns \cref{equ:psd_formula_real,equ:psd_formula_ghost} when there is no coupling, $k_\mathrm{r} = k_\mathrm{g} = 0$.
As shown in \cref{apd:fig:asymetry_settings}, the~model shows that asymmetric coupling strengths cause the discrepancy in peak heights at one resonant frequency, while asymmetric noise causes the dip in-between the resonant~frequencies.

If we are interested in the in-phase and out-of-phase modes $\{q_\mathrm{i}=q_\mathrm{r} + q_\mathrm{g},\ q_\mathrm{o}=q_\mathrm{r} - q_\mathrm{g}\}$, the~corresponding PSDs for these modes can be obtained by following the above steps and taking the transformation
\begin{equation}
    \mathbf{S}'_{qq} = \braket{\mathbf{T}\mathbf{q}(\omega) \mathbf{q}(\omega)^\dagger\mathbf{T}^\dagger}, \qquad \text{with } \mathbf{T} =
    \begin{pmatrix}
        1 & 1 \\
        1 & -1
    \end{pmatrix},
\end{equation}
which gives the corresponding {terms} 
\begin{subequations}
\begin{align}
        S_{qq}^{\prime\,\mathrm{i,i}} (\omega) &= S_{qq}^\mathrm{r,r} (\omega)+S_{qq}^\mathrm{g,g} (\omega)+S_{qq}^\mathrm{r,g} (\omega)+S_{qq}^\mathrm{g,r} (\omega), \\
        S_{qq}^{\prime\,\mathrm{o,o}} (\omega) &= S_{qq}^\mathrm{r,r} (\omega)+S_{qq}^\mathrm{g,g} (\omega)-S_{qq}^\mathrm{r,g} (\omega)-S_{qq}^\mathrm{g,r} (\omega), \\
        S_{qq}^{\prime\,\mathrm{i,o}} (\omega) &= S_{qq}^\mathrm{r,r} (\omega)-S_{qq}^\mathrm{g,g} (\omega)-S_{qq}^\mathrm{r,g} (\omega)+S_{qq}^\mathrm{g,r} (\omega), \\
        S_{qq}^{\prime\,\mathrm{o,i}} (\omega) &= S_{qq}^\mathrm{i,o} (\omega)^\ast.
    \end{align}
\end{subequations}
In the ideal scenario, where two coupled systems share the same properties (namely, $\Gamma_{r}=\Gamma_\mathrm{g}$, $\omega_{r}=\omega_\mathrm{g}$, 
$k_{r}=k_\mathrm{g}$, $M_\mathrm{r}=M_\mathrm{g}$ and $\xi_{r}=\xi_\mathrm{g}$), one has $S_{qq}^\mathrm{r,r}= S_{qq}^\mathrm{g,g}$, $S_{qq}^\mathrm{r,g}=S_{qq}^\mathrm{g,r}$ and thus $S_{qq}^{\prime\,\mathrm{i,o}}=0$, indicating that the in-phase and out-of-phase modes are the decoupled normal~modes.

\section{Quadratic mean of the coupling strength}\label{apd:sec:quadratic_mean_of_coupling}
\begin{figure}[tb]
\vspace{-0.4cm}
\begin{overpic}[width=\linewidth]{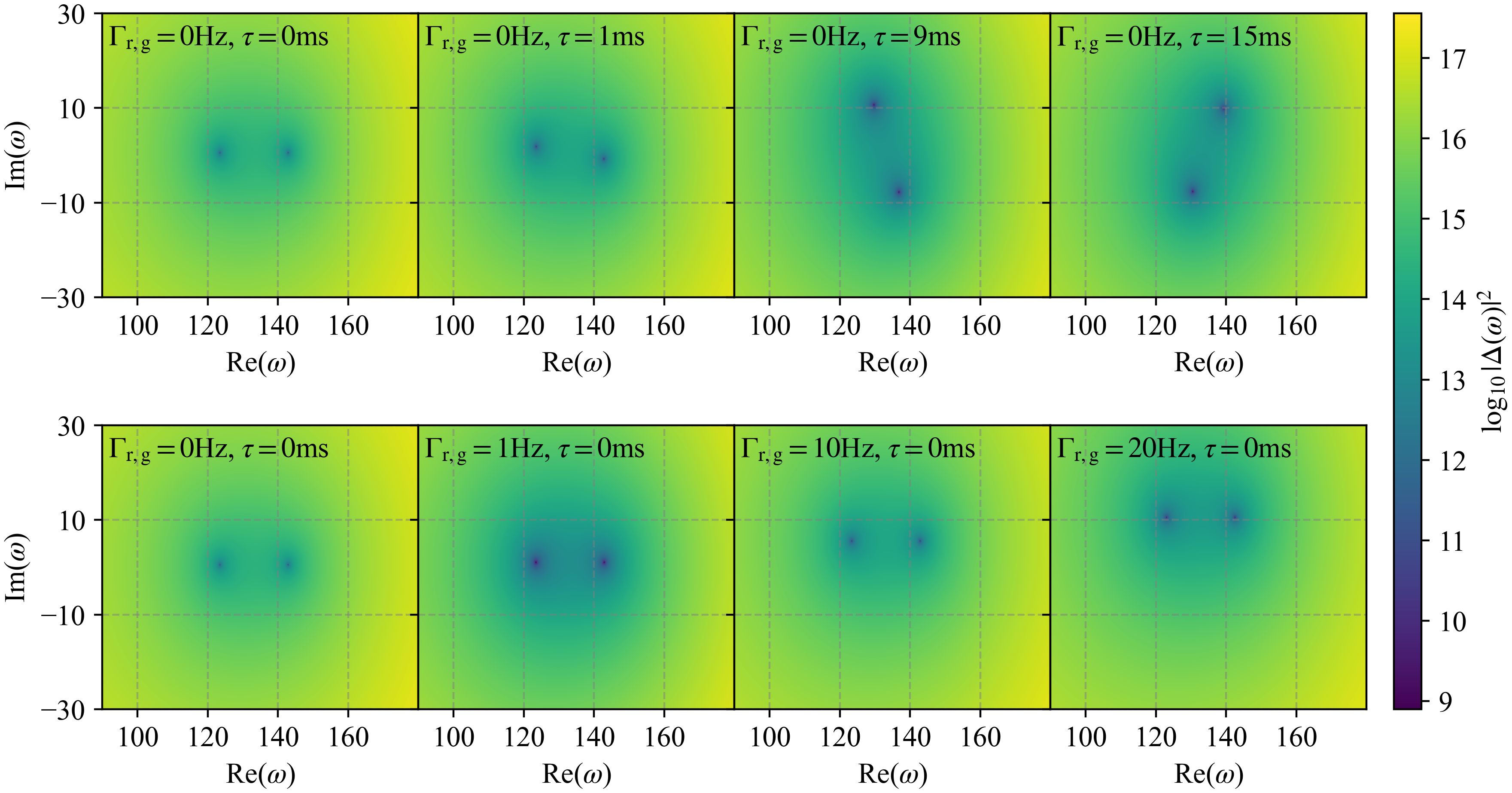}
    \put(0,51){(\textbf{a})} 
    \put(0,23){(\textbf{b})}
\end{overpic}
\caption{Numerical solutions of the resonant frequencies $\Delta(\omega)=0$, shown by the dark points, for~(\textbf{a}) different time delays $\tau$ and (\textbf{b}) different damping rates $\Gamma_\mathrm{r,g}$. Here, $\omega_\mathrm{r}=\omega_\mathrm{g}=2\pi\times123.5\,\unit{\Hz}$. \label{apd:fig:complex_omega_with_Gamma_tau}}
\end{figure} 

The average coupling strength $g$ in the coupled system described by \cref{equ:coupled_dynamics} can be inferred from the frequencies of coupled modes.
In general, these resonant frequencies obtained from $\Delta(\omega)=0$, given in Appendix~\ref{apd:sec:PSDs_for_coupled_systems}, are complex due to the presence of damping $\Gamma_\mathrm{r,g}$ and the time delay $\tau$, as~shown in \cref{apd:fig:complex_omega_with_Gamma_tau}. However, the~deviation of the solutions from the real axis is negligible for small damping and time delay. Under~the condition that $\omega_\mathrm{r}=\omega_\mathrm{g}=\omega_0$, in~the weak-damping regime $\Gamma_\mathrm{r,g}\ll\omega_\mathrm{r,g}$ and short time delay, we ignore their contributions (i.e.,~setting $\Gamma_\mathrm{r,g}=\tau=0)$ and write
\begin{equation}
    \Delta(\omega)\equiv \omega^4 - (\Omega_\mathrm{r}^2+\Omega_\mathrm{g}^2)\omega^2 + \left(\Omega_\mathrm{r}^2\Omega_\mathrm{g}^2 - \frac{k_\mathrm{r}k_\mathrm{g}}{M_\mathrm{r}M_\mathrm{g}}\right)=0,
\end{equation}
where we set $\Omega_\mathrm{r} = \omega_\mathrm{0}^2 +\frac{k_\mathrm{r}}{M_\mathrm{r}}$ and $\Omega_\mathrm{g} = \omega_\mathrm{0}^2 + \frac{k_\mathrm{g}}{M_\mathrm{g}}$. The~coupled frequencies are the solutions to this equation, which reads
\begin{equation}
    \omega_\mathrm{i,o}^2 = \frac{1}{2}(\Omega_\mathrm{r}^2+\Omega_\mathrm{g}^2) \pm \frac{1}{2}\sqrt{(\Omega_\mathrm{r}^2+\Omega_\mathrm{g}^2)^2-4\left(\Omega_\mathrm{r}^2\Omega_\mathrm{g}^2 - \frac{k_\mathrm{r}k_\mathrm{g}}{M_\mathrm{r}M_\mathrm{g}}\right)}.
\end{equation}
One can notice that, given the linear coupling in our model, the~difference between these two frequencies only depends on the coupling strengths, such that
\begin{align}
    \omega_\mathrm{o}^2-\omega_\mathrm{i}^2 &= \sqrt{(\Omega_\mathrm{r}^2+\Omega_\mathrm{g}^2)^2-4\left(\Omega_\mathrm{r}^2\Omega_\mathrm{g}^2 - \frac{k_\mathrm{r}k_\mathrm{g}}{M_\mathrm{r}M_\mathrm{g}}\right)}, \notag \\
    &=\sqrt{(\Omega_\mathrm{r}^2 - \Omega_\mathrm{g}^2)^2 + 4 \frac{k_\mathrm{r}k_\mathrm{g}}{M_\mathrm{r}M_\mathrm{g}}}, \notag \\
    &=\sqrt{\left(\frac{k_\mathrm{r}}{M_\mathrm{r}}+\frac{k_\mathrm{g}}{M_\mathrm{g}}\right)^2}, \notag \\
    &=\frac{k_\mathrm{r}}{M_\mathrm{r}}+\frac{k_\mathrm{g}}{M_\mathrm{g}},
\end{align}
with $\omega_\mathrm{i} = \omega_0$ and $\omega_\mathrm{0} = \sqrt{\omega_0^2 + \frac{k_\mathrm{r}}{M_\mathrm{r}}+\frac{k_\mathrm{g}}{M_\mathrm{g}}}$.
Thus, one can define the individual coupling rates $g_\mathrm{r}=\sqrt{k_\mathrm{r}/M_\mathrm{r}}$ and $g_\mathrm{g}=\sqrt{k_\mathrm{g}/M_\mathrm{g}}$, and~the quadratic mean coupling strength
\begin{equation}
    g=\sqrt{\frac{1}{2}\left(g_\mathrm{r}^2+g_\mathrm{g}^2\right)},
\end{equation}
which gives \cref{equ:coupling_rate_fomula} in the main~text.

\printbibliography

\end{document}